# First principles study of chalcogen vacancy effect on the optoelectronic and photocatalytic properties of transition metal dichalcogenides monolayers†

Tahir Wahab,[*a] Antonio Cammarata[*a] and Tomas Polcar [b]

In working conditions, chalcogen vacancies spontaneously occur in two-dimensional transition metal dichalcogenides (TMDCs) monolayers, affecting their optoelectronic and photocatalytic properties. To study how chalcogen vacancies affect such properties, we use quantum mechanical calculations considering prototypical $MX_2$ (M = Mo, W, X = S and Se) TMDCs monolayers. Structural optimisations show that M-X bond lengths about a vacancy are different compared to the bond lengths in the pristine structure. Band structure calculations reveal that the introduction of vacancies produce electronic states about the Fermi level, hence resulting in the reduction of the band gap. Work function and electrostatic potential calculations show that the introduction of vacancies induce an asymmetry in the electrostatic potential facilitating the charge separation; such feature is absent in a pristine monolayer. All the considered defective structures are capable of performing hydrogen evolution reaction, while co-catalyst is required to perform oxygen evolution reaction when used for water splitting. $WS_2$ and $WSe_2$ defective monolayers can serve as an efficient photocatalytic material for reducing $CO_2$ into useful chemical products. The presented results show that vacancy-containing TMDCs monolayers own photocatalytic capabilities compared to the pristine counterparts, thus showing that defective TMD monolayers have prospective applications and should not be regarded as flawed products to be discarded. Finally, the results might constitute guidelines for the experimental synthesis of vacancy-engineered $MX_2$ monolayers for optoelectronic devices and photocatalytic applications.

## 1 Introduction

Photocatalytic water splitting is an effective technique to convert solar light into molecular hydrogen and oxygen[1–3]. The basic principle of photocatalytic water splitting is as follows: when a photon with energy equal to or greater than the band gap $E_g$ of a photocatalytic semiconducting material is absorbed, electrons are excited from the valence band (VB) to the conduction band (CB); this generates a hole in the VB, and results in the formation of an exciton which is an electron-hole (e-h) pair[3–6]. If the electron and hole in the pair are separated, they can migrate to the surface of the semiconductor, and can be used to split water molecules to perform hydrogen evolution reaction (HER) in the CB and oxygen evolution reaction (OER) in the VB[3,6]. In order to perform the HER and OER, a semiconductor must satisfy the following conditions: *i*) $E_g$ must be greater than 1.23 eV and smaller than 3 eV, *ii*) the gap width between the valence band maximum (VBM) and conduction band minimum (CBM) must straddle the water oxidation/reduction (redox) potential, and *iii*) the semiconductor must exhibit excellent charge carrier separation and transport, i.e., low recombination rate and efficient charge carrier separation [4,7,8]. After the discovery of water splitting with $TiO_2$ by Fujishima et al.[9], researchers are now considering a variety of photocatalytic materials, including metal oxides[10,11], metal sulfides[12–14], metal car-bides[15,16], oxyhalides[17,18,18], oxynitrides[19,20], and recently transition metal dichalcogenides (TMDCs)[1,21–23].

[a] *Department of Control Engineering, Faculty of Electrical Engineering, Czech Technical University in Prague, Technicka 2, 16627 Prague 6, Czech Republic. Fax: +420 224 91 8646; Tel: +420 224 35 5713; E-mail: wahabtah@fel.cvut.cz, cammaant@fel.cvut.cz*
[b] *Engineering Materials & nCATS, FEE, University of Southampton, SO17 1BJ, Southampton, United Kingdom.*

† Electronic Supplementary Information (ESI) available: Optimized geometries. See DOI: 00.0000/00000000.

Beyond HER, the photocatalytic reduction of $CO_2$ into useful chemical products such as formic acid, formaldehyde, methanol, and methane is another technique to store sunlight in chemical bonds and mitigate the greenhouse emission[24–27]. Similar to photocatalytic water splitting, a semiconductor that reduce $CO_2$ to useful hydrocarbons should be low cost, highly stable, capable of absorbing visible light, excellent charge carrier separation and charge transfer[4,28,29]. Various material studied for the photoreduction of $CO_2$ includes metal oxides and mixed oxides[30,31], metal sulfides[32–34], polymeric materials[35], metal-organic framework (MOF)[36,37], and TMDCs[38–40].

TMDCs are materials with formula $MX_2$, where M is a transition metal (TM) while X is a chalcogen (S, Se and Te), where the TM is sandwitched between two X layers[41]. Because of the stability, large surface area, tunable and direct band gap nature, higher electrical conductivity, enhanced light absorption and strong excitonic effects, the two-dimensional group-VI TMDCs are excellent photocatalysts[23,42–44]. However, the high carrier recombination rate of photogenerated charge carrier, low light harvesting capacity and slow reaction kinetics are the main reasons - which hinder TMDCs to be efficient photocatalytic materials[45]. Various techniques such as mechanical exfoliation[46], chemical exfoliation[47], hydrothermal intercalation[48], two-step expansion and intercalation[49] and chemical vapour deposition (CVD)[46,50,51] are used to exfoliate TMDCs from their bulk counterpart. During the synthesis, defects such as chalcogen vacancies are inevitable[52]. It has been shown that pure basal planes are inactive in stable TMDCs, while anion vacancies can act as suitable active sites[53,54]. First-principles calculations and investigations on samples obtained from mechanical exfoliation and CVD deposition showed that the presence of chalcogen vacancies is favourable in $MoS_2$[46,55]. Anion vacancies strongly influence the structural and optoelectronic properties, significantly enhancing the photocatalytic water splitting activity in these materials[56–58]. The introduction of vacancies can then improve the photocatalytic ability by tuning the electronic band structure, increase the surface reactivity by increasing the number of active sites, enhance charge transport pathways, and influence the binding energy of all adsorbed species[56,59,60]. In this framework, we here investigate the effect of chalcogen vacancies on the optoelectronic and photocatalytic performance of prototypical $MX_2$ monolayers. The introduction of vacancies result in the reduction of the band gap by producing electronic states within the band gap. The optical absorption coefficient show a red shift in $MoX_2$ monolayer, while all the structures are capable of performing hydrogen evolution reaction, and photocatalytic reduction of $CO_2$ into useful products.

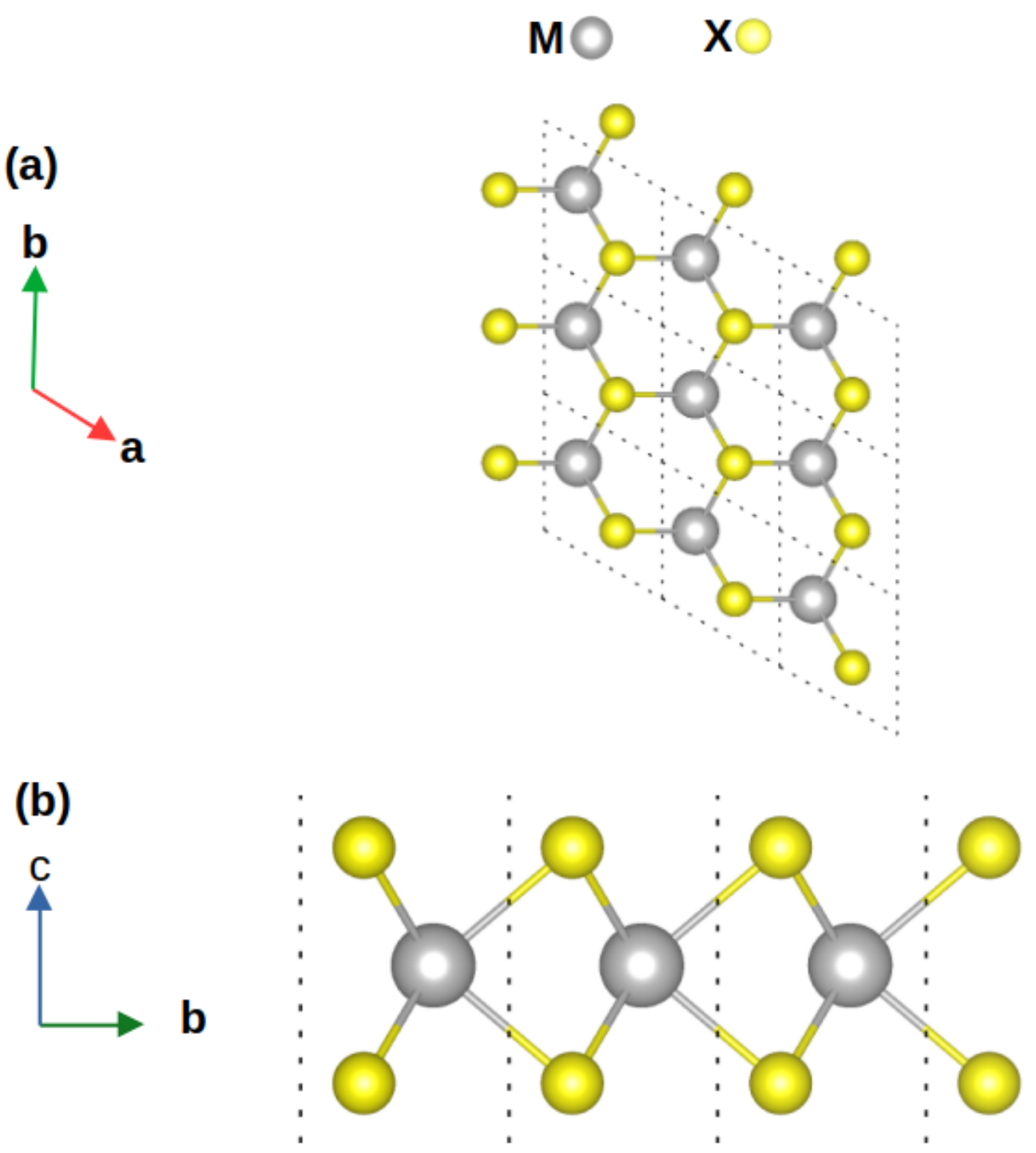

**Fig. 1** (a) Top and (b) side view of pristine $MX_2$ monolayer model geometry; the $\boldsymbol{a}, \boldsymbol{b}, \boldsymbol{c}$ lattice parameters are reported for reference. The dashed lines indicate the limits of the unit cell.

## 2 Calculation method

The reference for our computational models is the geometry of the prototypical $MX_2$ monolayer, with M = Mo, W and X = S and Se, which we already used to study the optical properties and photocatalytic activity of TMDC-based van der Waals heterostructures[61] (Figure 1). We use the cluster expansion approximation of the interactions as implemented in the Alloy Theoretic Automated Toolkit (ATAT)[62] coupled with the Vienna ab initio simulation package (VASP)[63] to find the vacancy concentration which realises the minimum of the energy convex hull. In this way, we identify the defective structure which is the most energetically favoured to form (Figure 2). We perform density functional theory calculations by using the Perdew-Burke-Ernzerhof (PBE)[64] version of the generalized gradient approximation. The lattice parameters of the defective structures are large enough (see Supplementary Information, section “Optimised geometries”) to justify the use of only the Γ-point to sample the Brillioun zone. The model geometries are relaxed by using a tolerance of $10^{-8}$ eV and $10^{-3}$ eV/Å on the self consistent field procedure and the forces, respectively. A convex hull[65] is calculated as a func-tion of the vacancy concentration for each selected $MX_2$ chemical composition; the structure realising the minimum in the convex hull is then selected for further investigation. All the first principles calculations are performed using the VASP software employing the projector-augmented wave[66] method, and using a minimum of 450 eV for the plane wave cutoff. To account for the long range van der Waals interactions, Grimme's (DFT-D2)[67] correction is used. To investigate the thermal stability of the selected structures, we perform ab initio molecular dynamics simulations[68] using Nosé-Hoover thermostat[69] with temperature equal to 300K with a time step of 1 fs and a total time window of 8 ps. The dynamical stability of all the structures is then examined by calculating the phonon dispersion by means of the PHONOPY software[70,71]. We calculate the electronic band structure of the selected geometries by using the PBE[72] and Heyd-Scuseria-Ernzerhof (HSE06) functionals[73]. Phonon dispersions and electronic band structures are calculated along the standard[74] piecewise linear path $\Gamma$-$M$-$K$-$\Gamma$ in the irreducible Brillioun zone. To examine the variation of the electrostatic potential along the $\boldsymbol{c}$-axis and the charge redistribution, we perform a planar average electrostatic potential calculations along the same axis at the HSE06 level.

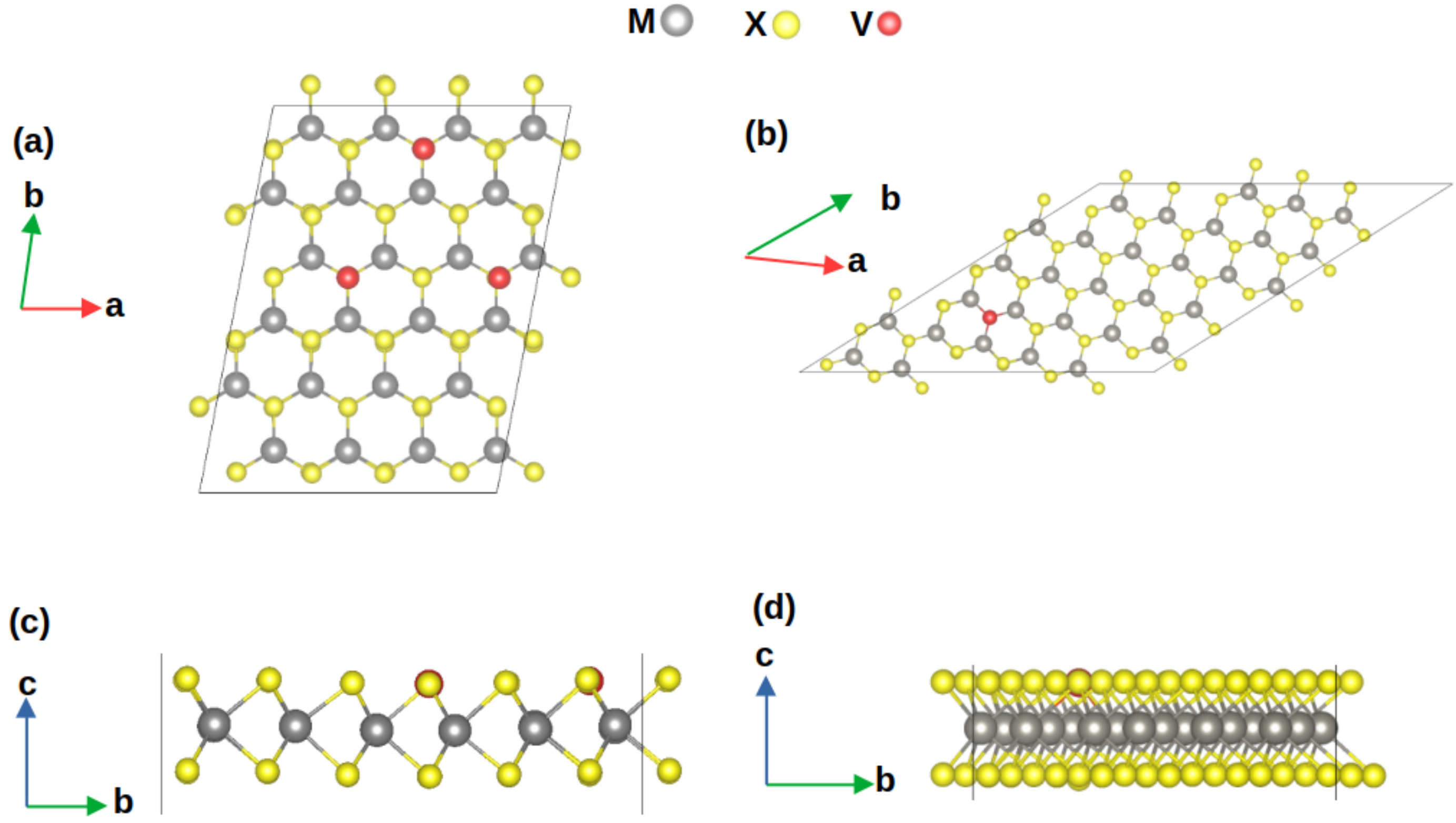

**Fig. 2** Example of defective $MX_2$ monolayer: (a-b) top and (c-d) side views showing the presence of a vacancy (indicated with a red sphere and labeled as "V") at the chalcogen site.

## 3 Results and Discussion

As mentioned in the previous section, we already investigated the optoelectronic properties and photocatalytic water spliting performance of pristine $MX_2$ monolayers[61]; we showed that all the monolayers straddle the water redox potential, and hence are capable of water splitting due to their favourable band gap and band alignment for hydrogen and oxygen evolution reaction. As we know that defects such as chalcogen vacancies are inevitable, the goal here is to examine the effect of chalcogen vacancies on the optoelectronic properties and the photocatalytic performance.

The first step is to determine which chalcogen vacancy concentration is preferred in $MX_2$ monolayers. To this aim, we use ATAT to comprehensively enumerate all the vacancy configurations on the chalcogen sublattice and construct the convex hull of the formation energy as a function of the vacancy concentration. The concentrations realising the lowest energy in the convex hull are found to be 6.25 % for $MoS_2$, $MoSe_2$, $WSe_2$ defective systems, respectively, while 1.56% for the $WS_2$ system. Each vacancy is surrounded by three cations coordinating X anions in their first coordination shell (Figure 2). We do not observe a significant change in the M–X bond lengths in the defective structures when compared to the pristine geometries. This suggests that no significant local strains are occurring about the vacancies and, correspondingly, the stability of the structure should not be affected.

**Table 1** Average M–X bond lengths with the corresponding standard deviation about X vacancies (Å). P and V indicate the values for the pristine and defective monolayer, respectively; standard deviation for $WS_2$ is zero as all the W–S bond lengths have equal values. No standard deviation is reported for the pristine monolayers because the M cation coordinates X neighbouring atoms in a regular polyhedron. The values of the pristine monolayers are taken from Ref [61].

| Structure | bond length P / V |
|---|---|
| $MoS_2$ | 2.41 / 2.40 ±0.01 |
| $MoSe_2$ | 2.54 / 2.53 ±0.01 |
| $WS_2$ | 2.41 / 2.41 ±0.00 |
| $WSe_2$ | 2.54 / 2.56 ±0.01 |

To assess the structural stability of the selected systems, we first calculate the phonon dispersion (Figure 3) along the standard path in the Brillouin zone (see section 2). The phonon band structure is free of imaginary frequencies except for a small pocket near the Γ point in the acoustic branch, which is known to be a numerical artifact in 2D transition metal dichalcogenides as mentioned by Fal'ko et al[75]. To confirm that such imaginary frequencies are an artifact and that the structure is dynamically stable at operating temperature, we perform AIMD simulations at 300 K. Figure 4, Figure 5 and Figure 6 show the variation in the total energy, temperature and the fluctuation in cation-anion M–X distance $\Delta d_{M-X}(t)$ as a function of the time, respectively. This fluctuation in bond length is defined as $\Delta d_{M-X}(t) = d_{M-X}(t) - < d_{M-X} >$, where $d_{M-X}(t)$ is the M-X bond length at time step t, and $< d_{M-X} >$ is the time average $M_X$ bond length over the entire steps. For all the systems, we observe that the total energy exhibits small oscillations about the average value and that the temperature fluctuates about the target value of 300 K. At the same time, we see that the fluctuation in M–X bond length oscillates about the mean value over the entire trajectory, which indicates that no bond breaking occurs and the structure is thermodynamically stable. Therefore, phonon band and AIMD simulations confirm that the defective structures containing vacancies are dynamically and thermally stable, and it is then worthy to investigate their optical and photocatalytic response.

We then continue our analysis by calculating the electronic band structure at the PBE (Figure 7) and HSE06 (Figure 8) theory level. According to the PBE calculations, only $WSe_2$ exhibits a direct band gap, both VBM and CBM lying at the $K$ point, while all the remaining structures exhibit indirect band gap. It can be noted that (Figure 7) the introduction of chalcogen vacancies introduce states within the band gap which are not present in the pristine monolayer; as a result, the band gap of the defective structures is smaller compared to the gap width found in the pristine ones. According to HSE06 calculations (Figure 8), $WS_2$ and $WSe_2$ exhibit a direct band gap while $MoS_2$ and $MoSe_2$ exhibit indirect band gap. The HSE06 calculated band gap values for $MoS_2$, $MoSe_2$, $WS_2$, and $WSe_2$ are 1.25, 1.31, 1.85, and 1.53 eV respectively. We find that (Figure 8) the major contribution to both VBM and CBM comes from the transition metal $d$-orbitals[76] around the vacancy, while the chalcogen atom near the vacancy does not contribute to the energy levels about the band gap in a significant way. However we can also see some minor contribution by the chalcogen: in the case of $MoS_2$ and $MoSe_2$, the chalcogen has a minor contribution to VBM, while in the case of $WS_2$ it has a minor contribution to CBM.

In order to calculate the electrostatic potential asymmetry induced by the presence of vacancies, we calculate the planar average electrostatic potential along the $\boldsymbol{c}$-axis (Figure 9). As compared to the layer bearing vacancies, the electrostatic potential is more negative on the pristine chalcogen layer; this can be ascribed to the fact that the introduction of vacancies at the chalcogen site results in the reduction of local electron density and hence in a less negative electrostatic potential on the vacancy side. We then calculate the work function $\phi$ as

$$\phi = E_{vac} - E_F \tag{1}$$

where $E_F$ and $E_{vac}$ represent the Fermi and the vacuum level, respectively, the latter obtained from the plateau of the electrostatic potential in correspondence to the vacuum slab. The vacuum level on the vacancy side is larger as compared to the pristine anion side; this produces two different work function values, one for each side of the monolayer[61,77]. In all the structures, $\phi$ on the vacancy side is higher than the value on the pristine side, with a difference $\delta\phi$ of 21, 83, 19 and 89 meV for $MoS_2$, $MoSe_2$, $WS_2$, and $WSe_2$, respectively. It can be seen that $\delta\phi$ in the Se based structure is larger as compared to the S based structures; this correlates with the greater polarizability of Se compared to S, and hence a prominent asymmetry in the surface potential. In a pristine $MX_2$ monolayer, due to the symmetrical distribution of the chalcogen atoms with respect to the cation layer, the electrostatic potential is sym-metric with respect to the same layer; as a consequence, $\phi$ assumes the same value on both sides of the monolayer and there is no built-in electric field which can separate the photogenerated charge carriers. Instead, we can see that the introduction of vacancies on one of the two X layers results in an asymmetry of the electrostatic potential together with a $\delta\phi$, which determines a built-in electric field that facilitates charge separation and hence reduces charge recombination. Compared to $MS_2$ the larger values of $\delta\phi$ in $MSe_2$ suggest a strong built-in electric field to separate the photogenerated charge carriers and hence can increase the photocatalytic water splitting efficiency. The trends in the $\phi$, $\delta\phi$ and the electrostatic potentials is similar to what has been observed in a previous work[76], where the difference of the reported values is due to the different concentration of vacancies we have in the present work.

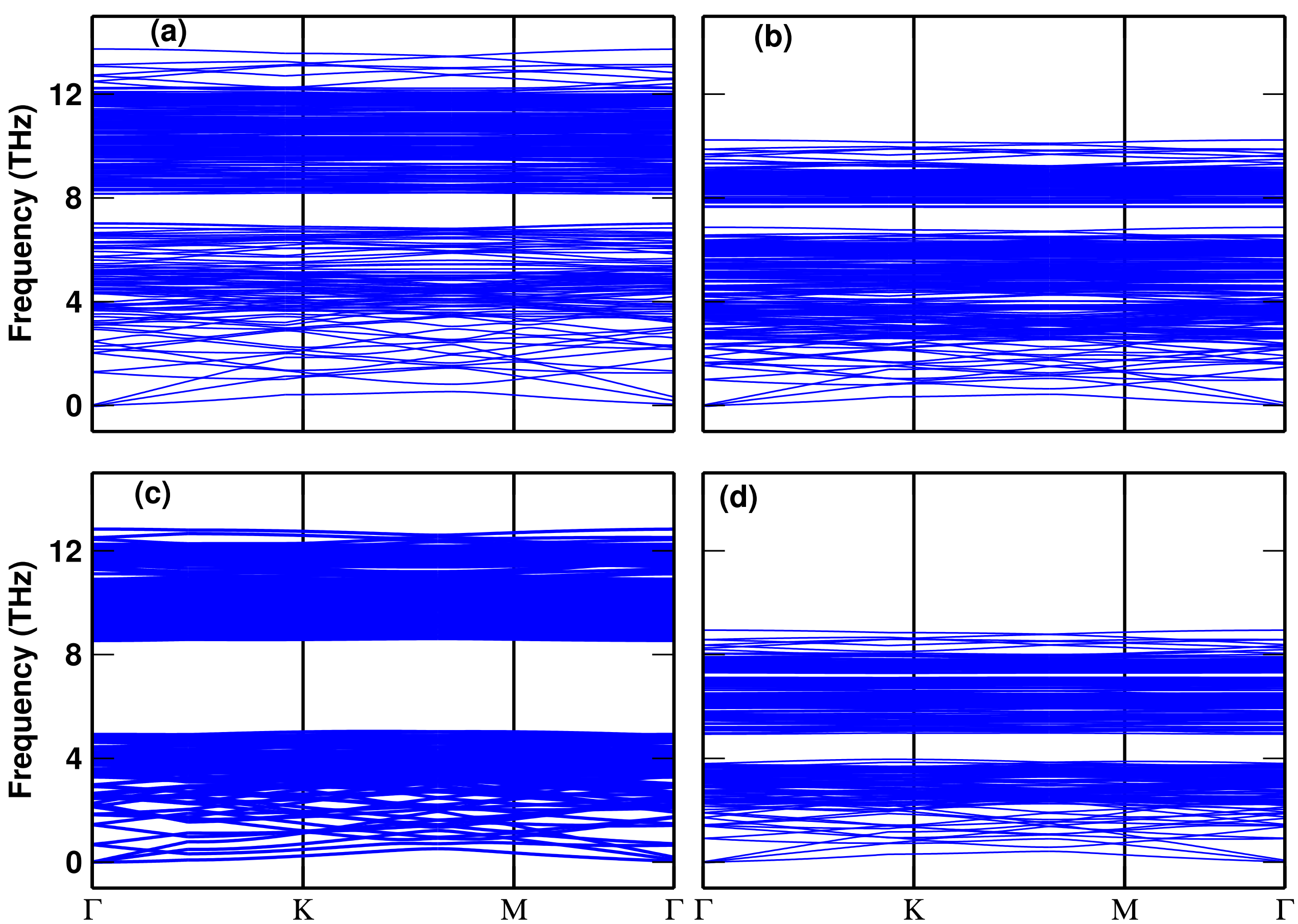

**Fig. 3** Phonon band dispersion along a piecewise linear path in the irreducible Brillioun zone of (a) $MoS_2$, (b) $MoSe_2$, (c) $WS_2$, and (d) $WSe_2$ systems. The dashed orange horizontal line indicates marks the position of zero frequency.

We then calculate the optical absorption coefficient $\alpha$[78] in the $[0,6]$ eV energy range (Figure 10). For $MoS_2$, the first absorption peak lies within the infrared region, while for $MoSe_2$ the first absorption peak lies near the visible region. Concerning $WS_2$, the absorption starts in the visible region, while the first absorption peak lies in the UV range, while for $WSe_2$ the first absorption peak lies within the visible range. It can be seen that, as compared to the pristine case, all the $MoX_2$ monolayers containing vacancies show a red shift in the first absorption peak[61]. In fact, as observed above, the presence of anion vacancies results in the introduction of electronic states within the band gap (Figure 8), which facilitates the absorption of light with lower energy than that required in the case of pristine monolayers[61], including infrared radiation.

In order to check the photocatalytic water splitting activity, we compare the HSE06 calculated VBM and CBM with water redox potential (Figure 11 (a)). We here recall that a semiconductor must satisfy two conditions to split the water molecule: the band gap should be greater than 1.23 eV, and the band edges must straddle the water redox potential[79]. The HSE06 calculated band gap value of all the considered structure is greater than 1.23 eV. Also, it can be seen that the CBM of all the monolayers lie above the potential required for HER, suggesting that, under visible light illumination, the electrons in the CBM have enough energy to reduce hydrogen from the water molecule. However, the VBM also lies above the potential required for the OER, this suggesting a limited capability of performing OER under sun light.

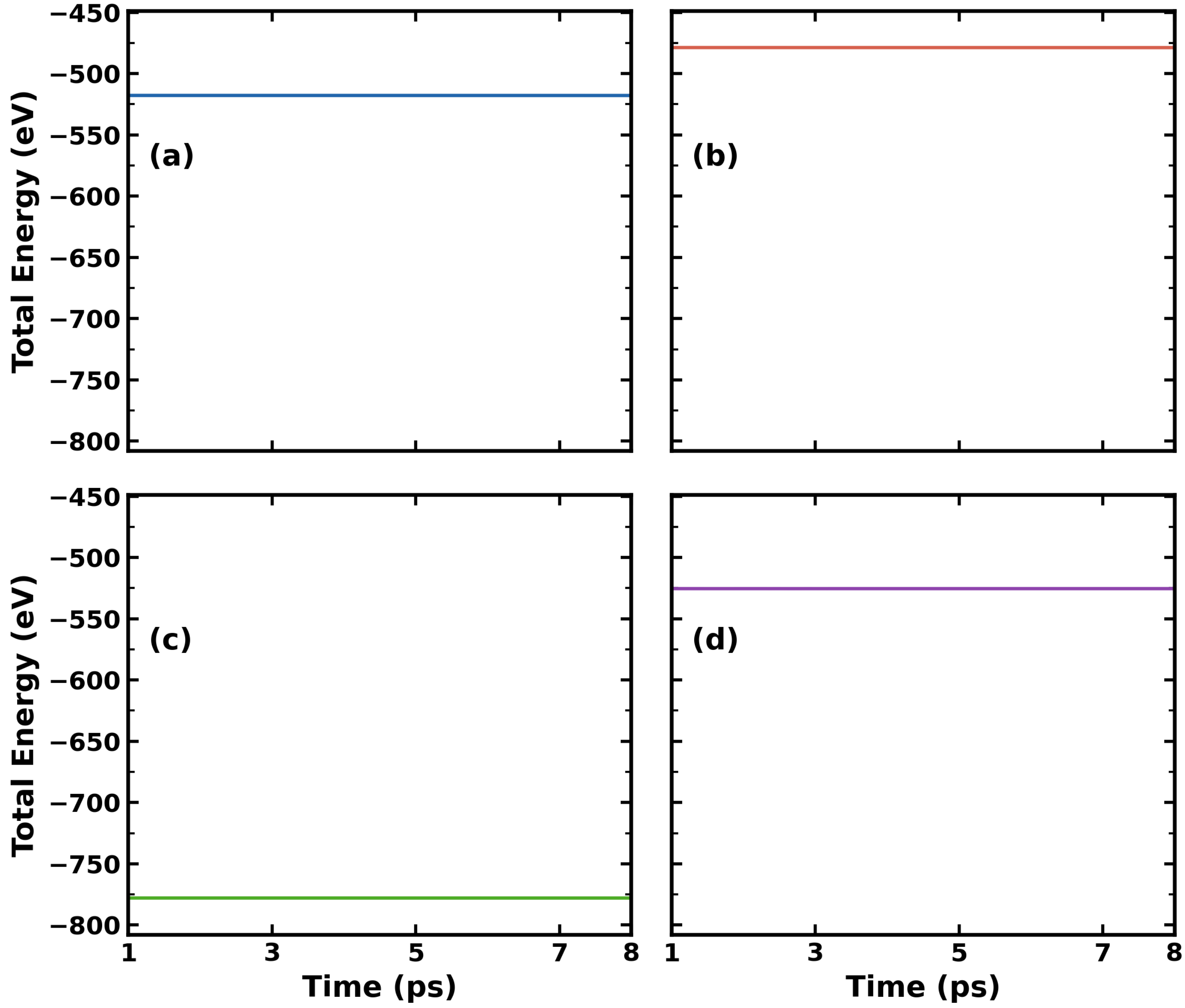

**Fig. 4** Total energy as a function of the simulation time of the (a) $MoS_2$, (b) $MoSe_2$, (c) $WS_2$, and (d) $WSe_2$ systems.

Next we examine the the $CO_2$ reduction capability of these structures by comparing the HSE06 calculated CBM with $CO_2$ reduction products on the vacuum scale (Figure 11 (b)). The equations representing the reduction half reactions and converting $CO_2$ in carbon products[30,80,81] are as follows

$$CO_2 + 2H^+ + 2e^- \rightarrow HCOOH \quad (2)$$

$$CO_2 + 2H^+ + 2e^- \rightarrow CO + H_2O \quad (3)$$

$$CO_2 + 4H^+ + 4e^- \rightarrow HCHO + H_2O \quad (4)$$

$$CO_2 + 6H^+ + 6e^- \rightarrow CH_3OH + H_2O \quad (5)$$

$$CO_2 + 8H^+ + 8e^- \rightarrow CH_4 + 2H_2O. \quad (6)$$

At pH = 7 the potential required for the half reduction reaction to convert $CO_2$ to formic acid (HCOOH), carbon monoxie (CO), formaldehyde (HCHO), methanol ($CH_3OH$) and methane ($CH_4$) are -3.83, -3.91, -3.96, -4.06 and -4.20 eV with respect to the vacuum level respectively, while the potential requires for OER is -5.26[81]. We know that the minimum band gap requirement for a semiconductor to split water is 1.23 eV, which is the difference of potential between HER and OER. As in case of $CO_2$ reduction, we can have many products (from formic acid to methane), thus the band gap requirement is product dependent. According to Equations (2)–(6), the thermodynamic potential difference between the oxidation potential and the corresponding reduction potential of $CO_2$ are 1.43, 1.35, 1.30, 1.20, and 1.06 eV respectively. It can be seen that the CBM of all the studied systems lies above the potential required to carry out the half reduction reaction. The HSE06 calculated band gap values show that: *i*) $WS_2$ and $WSe_2$ are capable of reducing $CO_2$ into all the products, *ii*) $MoSe_2$ is capable of reducing $CO_2$ into HCHO, $CH_3OH$, and $CH_4$, *iii*) $MoS_2$ is capable of reducing $CO_2$ into methanol and methane. Thus, among the considered systems, $WS_2$ and $WSe_2$ are the most promising candidates for the photocatalytic reduction of $CO_2$ and are capable of performing HER in water splitting, while a co-catalyst would be required to assist the OER.

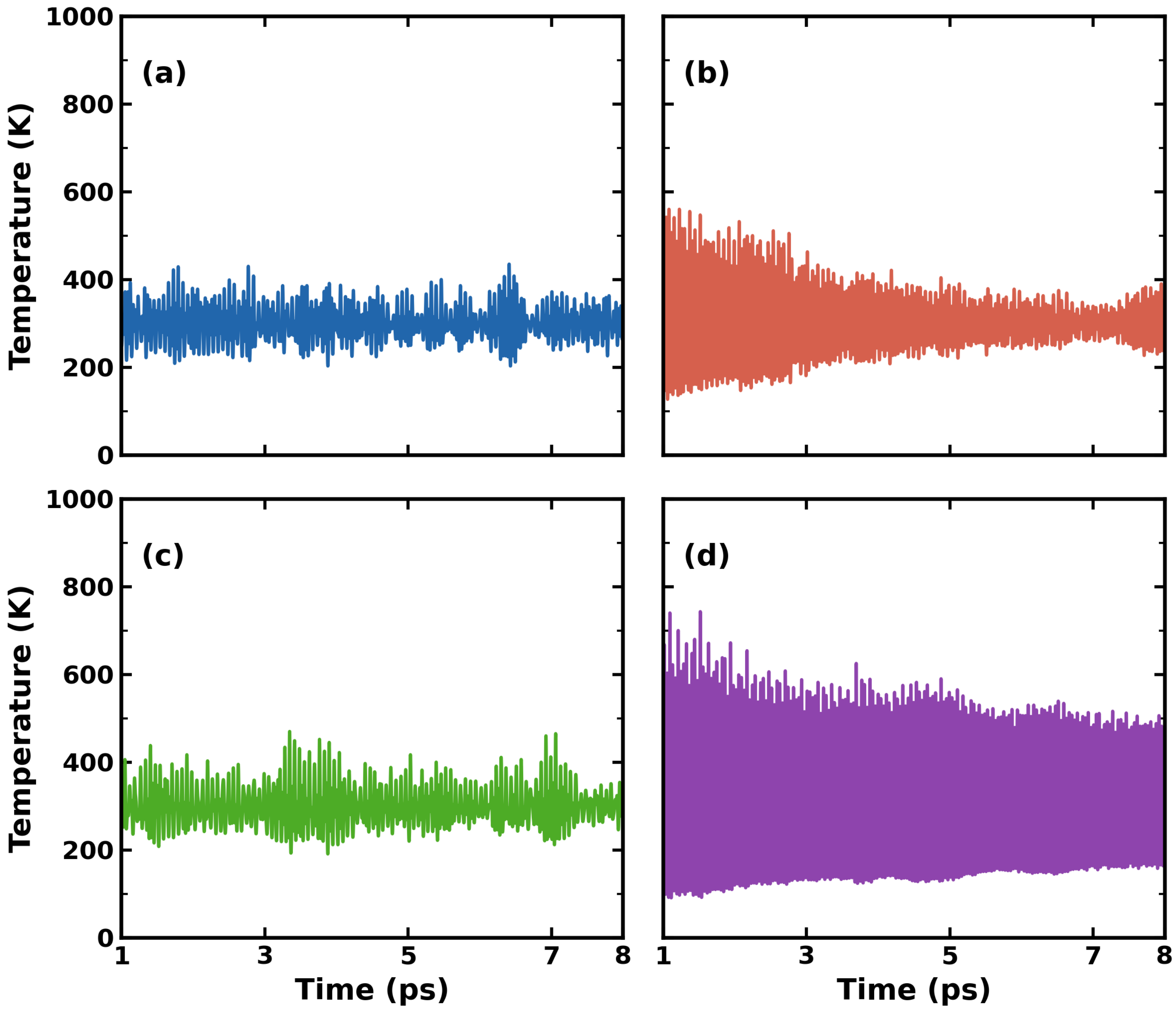

**Fig. 5** Temperature as a function of the simulation time of the (a) $MoS_2$, (b) $MoSe_2$, (c) $WS_2$, and (d) $WSe_2$ systems.

## 4 Conclusions

We have systematically examined the effect of chalcogen vacancies on the structural, optoelectronic properties and photocatalytic water splitting activity in $MX_2$ monolayers.

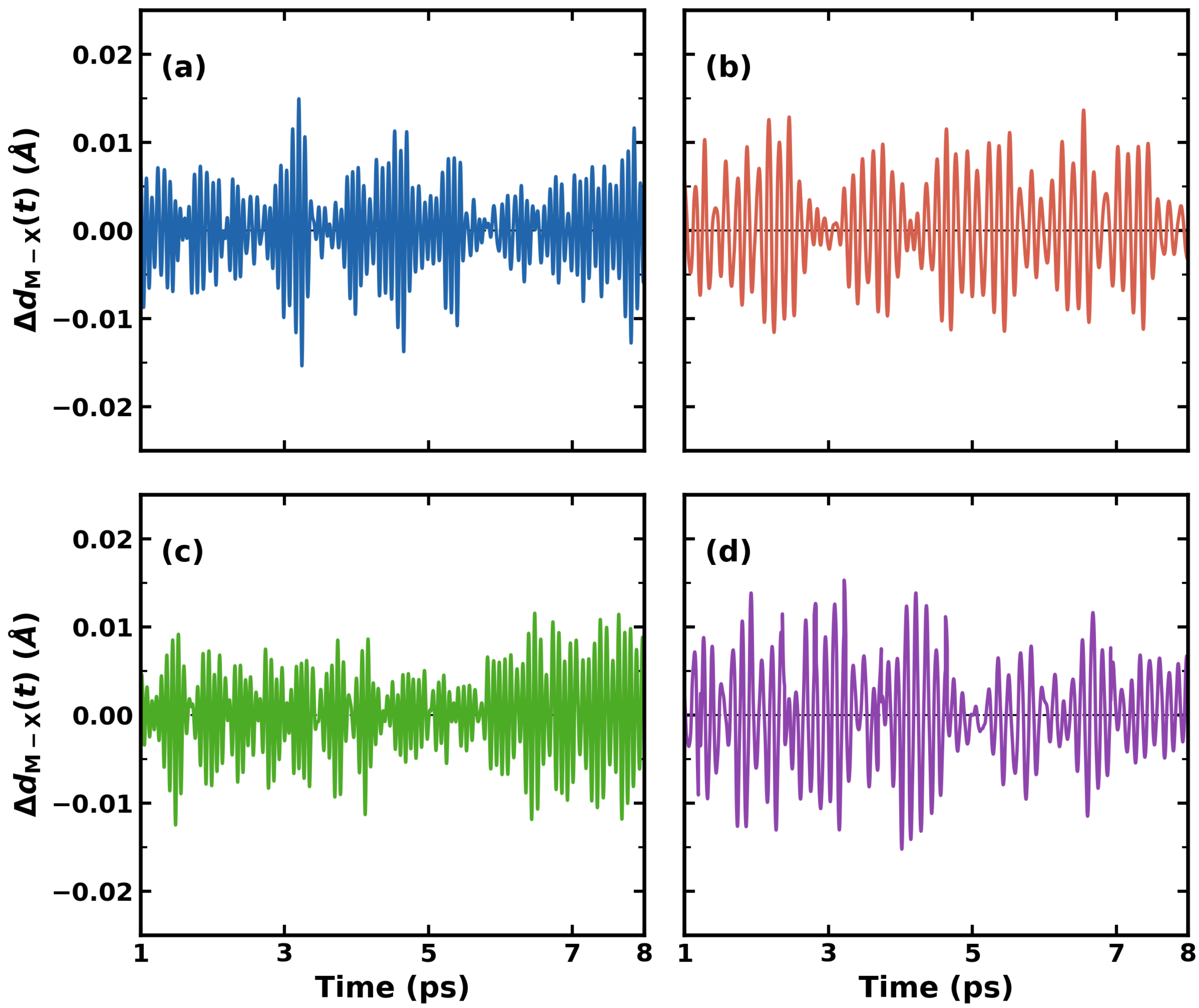

**Fig. 6** M–X bond length as a function of the simulation time of the (a) $MoS_2$, (b) $MoSe_2$, (c) $WS_2$, and (d) $WSe_2$ systems.

Vacancy concentrations which is likely to form in $MoS_2$, $MoSe_2$, and $WSe_2$ monolayer is 6.25%, while for $WSe_2$ it is 1.56%; such concentrations have been considered for the present study. The structural analysis revealed that M–X bond lengths about the vacancies are very close to those found in the pristine structure, this suggesting the absence of local strains which might affect the stability of the atom geometry. The thermal and dynamical stability of all the structures has been examined by performing ab initio molecular dynamic simulations and phonon band structure calculations, which showed that all the structures are thermally and dynamically stable. The introduction of vacancies resulted in the introduction of electronic states within the band gap, with a consequent reduction of the gap width. The calculated projected band structures showed that the major contribution to both VBM and CBM come from the $d$–orbitals of the transition metal, while some minor contribution from the chalcogen atoms can also be observed in some cases. The planar average electrostatic potential analysis revealed that there is a difference of potential between the X layer bearing vacancies and the pristine X layer, with a work function higher on the vacancy side; this favours charge separation across the layer. Optical absorption coefficient calculations showed that introduction of vacancies resulted in a red-shift of the first absorption peak, while the calculated band edge positions showed that the monolayers are capable of performing hydrogen evolution reaction. Among the considered structures, only $WS_2$ and $WSe_2$ satisfy the conditions of reducing $CO_2$ into useful products, due to favorable conduction band edge positions and band gap which can absorb visible light efficiently; at the same time, all the stuctures are capable of performing HER in water splitting, while a co-catalyst would be required to assite the OER. These results might constitute guidelines for the experimental synthesis of vacancy-engineered $MX_2$ monolayers for optoelectronic devices and photocatalytic applications.

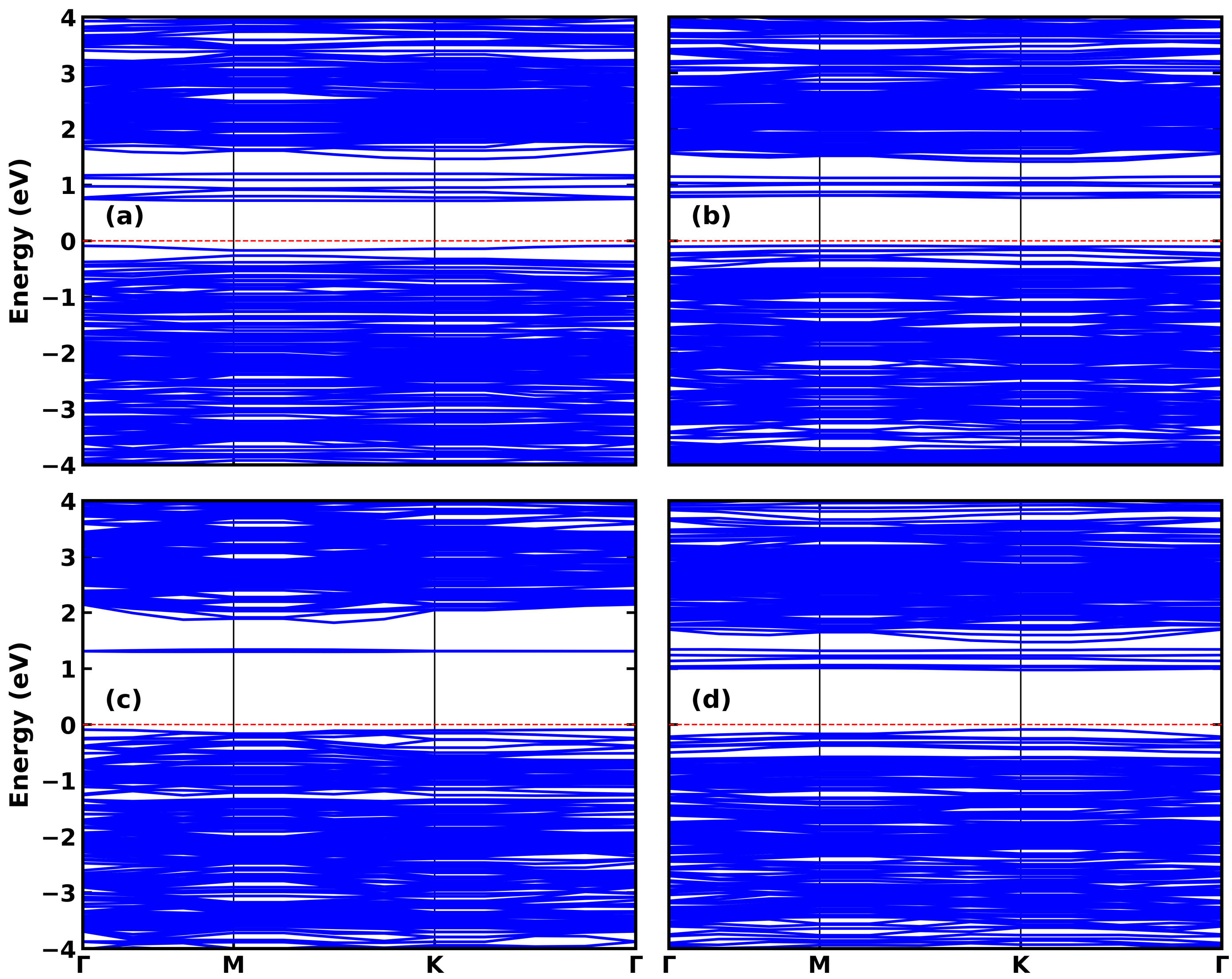

**Fig. 7** Electronic band structure of (a) $MoS_2$, (b) $MoSe_2$, (c) $WS_2$, and (d) $WSe_2$ defective systems calculated along the high symmetry path by using the PBE energy functional. The Fermi level has been set to 0 eV and is indicated by a dashed line.

## Conflicts of interest

There are no conflicts to declare.

## Acknowledgements

This work was co-funded by the European Union under the project "Robotics and advanced industrial production" (reg. no. CZ.02.01.01/00/22_008/0004590). This work was also done with the support of the Czech Science Foundation (project No. 24-12643L), and by the Ministry of Education, Youth and Sports of the Czech Republic through the e-INFRA CZ (ID:90254). The access to the computational infrastructure of the OP VVV funded project CZ.02.1.01/0.0/0.0/16_019/0000765 "Research Center for Informatics" and the use of the VESTA software[82] is also acknowledged.

## Supplementary Information

## Notes and references

1 V. Takhar and R. Banerjee, *ChemCatChem*, 2025, **17**, e00405.

2 I. Ahmad, I. Shahid, A. Ali, E. Li, Y. Gan, X. Zhu, X. Chen, X.-P. Wang and F. Rao, *International Journal of Hydrogen Energy*, 2025, **128**, 523–533.

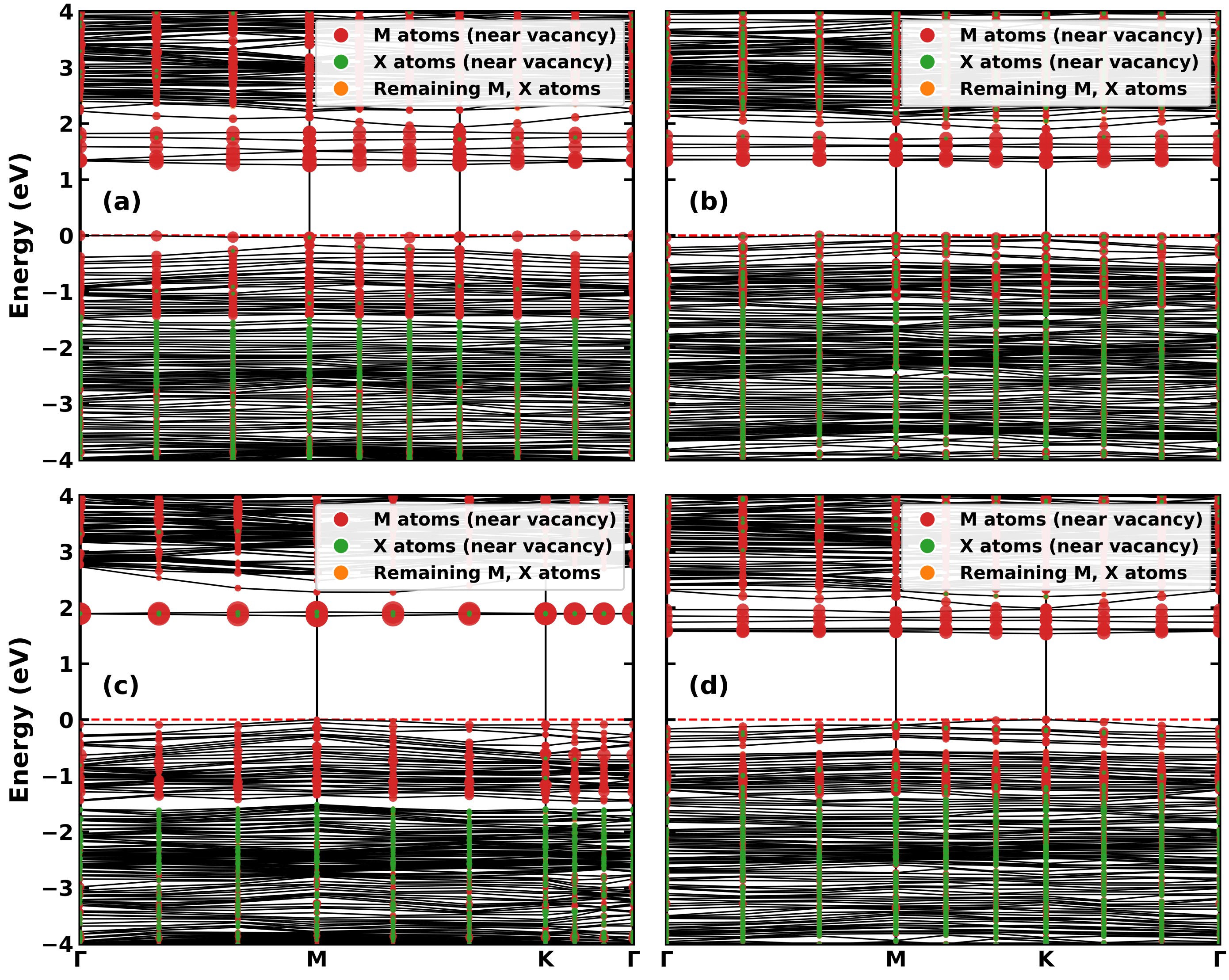

**Fig. 8** Atom-resolved band structure of (a) $MoS_2$, (b) $MoSe_2$, (c) $WS_2$, and (d) $WSe_2$ defective systems, calculated at the HSE06 theory level. The Fermi level has been set to 0 eV and is indicated by a dashed line.

3 A. O. M. Almayyali, M. Merdan, H. R. Jappor and S. J. Musa, *Journal of Physics and Chemistry of Solids*, 2025, **207**, 112997.

4 A. K. Singh, K. Mathew, H. L. Zhuang and R. G. Hennig, *The journal of physical chemistry letters*, 2015, **6**, 1087–1098.

5 H. L. Zhuang and R. G. Hennig, *Chemistry of Materials*, 2013, **25**, 3232–3238.

6 A. Wu, Y. Xie, H. Ma, C. Tian, Y. Gu, H. Yan, X. Zhang, G. Yang and H. Fu, *Nano Energy*, 2018, **44**, 353–363.

7 S. Cao and L. Piao, *Angewandte Chemie International Edition*, 2020, **59**, 18312–18320.

8 C.-F. Fu, X. Wu and J. Yang, *Advanced materials*, 2018, **30**, 1802106.

9 A. Fujishima and K. Honda, *nature*, 1972, **238**, 37–38.

10 M. M. Khan, S. F. Adil and A. Al-Mayouf, *Metal oxides as photocatalysts*, 2015.

11 T. Velempini, E. Prabakaran and K. Pillay, *Materials Today Chemistry*, 2021, **19**, 100380.

12 L. Jia, X. Tan, T. Yu and J. Ye, *Energy & Fuels*, 2022, **36**, 11308–11322.

13 Q. Zhu, Q. Xu, M. Du, X. Zeng, G. Zhong, B. Qiu and J. Zhang, *Advanced Materials*, 2022, **34**, 2202929.

14 S. Chandrasekaran, L. Yao, L. Deng, C. Bowen, Y. Zhang, S. Chen, Z. Lin, F. Peng and P. Zhang, *Chemical Society Reviews*, 2019, **48**, 4178–4280.

15 J. Fan, H. Wang, W. Sun, H. Duan and J. Jiang, *Materials Today*, 2024, **76**, 110–135.

16 A. Sinopoli, Z. Othman, K. Rasool and K. A. Mahmoud, *Current Opinion in Solid State and Materials Science*,

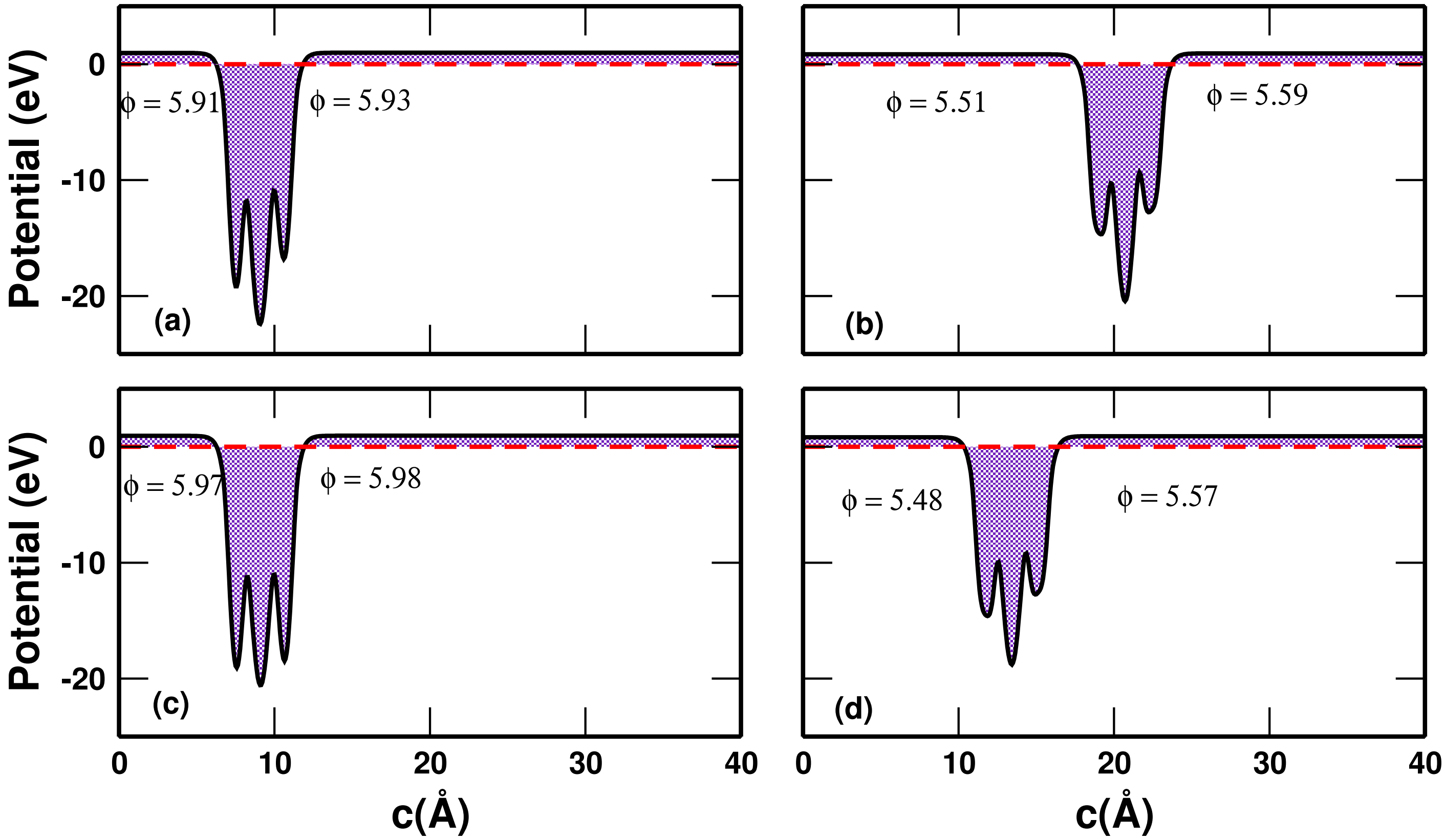

**Fig. 9** Planar-average electrostatic potential along the $c$ axis of (a) $MoS_2$, (b) $MoSe_2$, (c) $WS_2$, and (d) $WSe_2$ defective systems, calculated at the HSE06 theory level.

2019, **23**, 100760.

17 R. O. Gembo, R. Ratshiedana, L. M. Madikizela, I. Kamika, C. K. King'ondu, A. T. Kuvarega and T. A. Msagati, *Catalysis Science & Technology*, 2024, **14**, 6466–6495.

18 K. Ogawa, A. Nakada, H. Suzuki, O. Tomita, M. Higashi, A. Saeki, H. Kageyama and R. Abe, *ACS applied materials & interfaces*, 2018, **11**, 5642–5650.

19 M. Ahmed and G. Xinxin, *Inorganic Chemistry Frontiers*, 2016, **3**, 578–590.

20 A. Fuertes, *Materials Horizons*, 2015, **2**, 453–461.

21 K. Pramoda and C. Rao, *ChemNanoMat*, 2022, **8**, e202200153.

22 N. N. Rosman, R. M. Yunus, L. J. Minggu, K. Arifin, M. N. I. Salehmin, M. A. Mohamed and M. B. Kassim, *International Journal of Hydrogen Energy*, 2018, **43**, 18925–18945.

23 B. Peng, P. K. Ang and K. P. Loh, *Nano Today*, 2015, **10**, 128–137.

24 X. Nie, M. R. Esopi, M. J. Janik and A. Asthagiri, *Angewandte Chemie International Edition*, 2013, **52**, 2459–2462.

25 M. Gattrell, N. Gupta and A. Co, *Journal of electroanalytical Chemistry*, 2006, **594**, 1–19.

26 M. D. Porosoff, B. Yan and J. G. Chen, *Energy & Environmental Science*, 2016, **9**, 62–73.

27 Z. Weng, J. Jiang, Y. Wu, Z. Wu, X. Guo, K. L. Materna, W. Liu, V. S. Batista, G. W. Brudvig and H. Wang, *Journal of the American Chemical Society*, 2016, **138**, 8076–8079.

28 Ş. Neaţu, J. A. Maciá-Agulló and H. Garcia, *International journal of molecular sciences*, 2014, **15**, 5246–5262.

29 Z. Kovacic, B. Likozar and M. Hus, *ACS catalysis*, 2020, **10**, 14984–15007.

30 S. N. Habisreutinger, L. Schmidt-Mende and J. K. Stolarczyk, *Angewandte Chemie International Edition*, 2013, **52**, 7372–7408.

31 K. Teramura, T. Tanaka, H. Ishikawa, Y. Kohno and T. Funabiki, *The Journal of Physical Chemistry B*, 2004, **108**, 346–354.

32 S. Navalón, A. Dhakshinamoorthy, M. Álvaro and H. Garcia, *ChemSusChem*, 2013, **6**, 562–577.

33 H. Fujiwara, H. Hosokawa, K. Murakoshi, Y. Wada, S. Yanagida, T. Okada and H. Kobayashi, *The Journal of Physical Chemistry B*, 1997, **101**, 8270–8278.

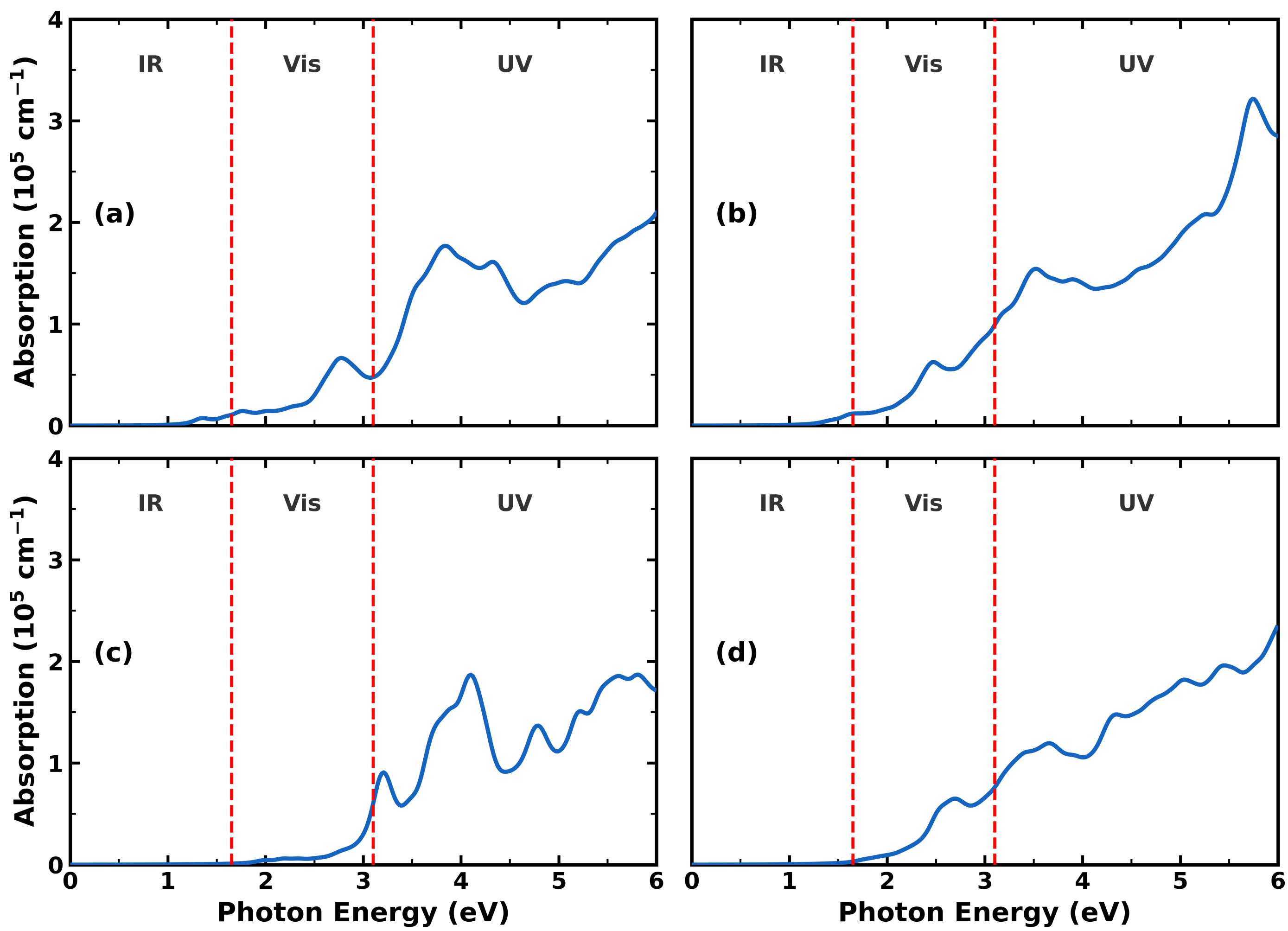

**Fig. 10** Calculated optical absorption spectrum of (a) $MoS_2$, (b) $MoSe_2$, (c) $WS_2$, and (d) $WSe_2$ defective systems.

34 S. Kuwabata, K. Nishida, R. Tsuda, H. Inoue and H. Yoneyama, *Journal of the Electrochemical Society*, 1994, **141**, 1498–1503.

35 R. Kuriki, K. Sekizawa, O. Ishitani and K. Maeda, *Angewandte Chemie International Edition*, 2015, **54**, 2406–2409.

36 S. Wang, W. Yao, J. Lin, Z. Ding and X. Wang, *Angewandte Chemie*, 2014, **126**, 1052–1056.

37 D. Sun, Y. Fu, W. Liu, L. Ye, D. Wang, L. Yang, X. Fu and Z. Li, *Chemistry–A European Journal*, 2013, **19**, 14279–14285.

38 Z. Li, N. H. Attanayake, J. L. Blackburn and E. M. Miller, *Energy & Environmental Science*, 2021, **14**, 6242–6286.

39 M. Asadi, K. Kim, C. Liu, A. V. Addepalli, P. Abbasi, P. Yasaei, P. Phillips, A. Behranginia, J. M. Cerrato, R. Haasch *et al.*, *Science*, 2016, **353**, 467–470.

40 Y. Zhao, Y. Chen, P. Ou and J. Song, *ACS catalysis*, 2023, **13**, 12941–12951.

41 Y. Jing, B. Liu, X. Zhu, F. Ouyang, J. Sun and Y. Zhou, *Nanophotonics*, 2020, **9**, 1675–1694.

42 R. Yang, Y. Fan, Y. Zhang, L. Mei, R. Zhu, J. Qin, J. Hu, Z. Chen, Y. Hau Ng, D. Voiry *et al.*, *Angewandte Chemie*, 2023, **135**, e202218016.

43 S. Manzeli, D. Ovchinnikov, D. Pasquier, O. V. Yazyev and A. Kis, *Nature Reviews Materials*, 2017, **2**, 17033.

44 K. F. Mak and J. Shan, *Nature Photonics*, 2016, **10**, 216–226.

45 M. Ai, J.-W. Zhang, Y.-W. Wu, L. Pan, C. Shi and J.-J. Zou, *Chemistry–An Asian Journal*, 2020, **15**, 3599–3619.

46 J. Hong, Z. Hu, M. Probert, K. Li, D. Lv, X. Yang, L. Gu, N. Mao, Q. Feng, L. Xie *et al.*, *Nature communications*, 2015, **6**, 6293.

47 H. Lin, J. Wang, Q. Luo, H. Peng, C. Luo, R. Qi, R. Huang, J. Travas-Sejdic and C.-G. Duan, *Journal of Alloys and Compounds*, 2017, **699**, 222–229.

48 M. Piao, J. Chu, X. Wang, Y. Chi, H. Zhang, C. Li, H. Shi and M.-K. Joo, *Nanotechnology*, 2018, **29**, 025705.

49 H. Li, Y. Shi, M.-H. Chiu and L.-J. Li, *Nano Energy*, 2015, **18**, 293–305.
50 L. Tang, T. Li, Y. Luo, S. Feng, Z. Cai, H. Zhang, B. Liu and H.-M. Cheng, *ACS nano*, 2020, **14**, 4646–4653.
51 S. L. Wong, H. Liu and D. Chi, *Progress in Crystal Growth and Characterization of Materials*, 2016, **62**, 9–28.
52 H. Qiu, T. Xu, Z. Wang, W. Ren, H. Nan, Z. Ni, Q. Chen, S. Yuan, F. Miao, F. Song *et al.*, *Nature communications*, 2013, **4**, 2642.
53 Y. Ji, J. K. Nørskov and K. Chan, *The Journal of Physical Chemistry C*, 2019, **123**, 4256–4261.
54 J. Lee, S. Kang, K. Yim, K. Y. Kim, H. W. Jang, Y. Kang and S. Han, *The Journal of Physical Chemistry Letters*, 2018, **9**, 2049–2055.
55 P. Vancsó, G. Z. Magda, J. Pető, J.-Y. Noh, Y.-S. Kim, C. Hwang, L. P. Biró and L. Tapasztó, *Scientific reports*, 2016, **6**, 29726.
56 Y. Jiang, H. Sun, J. Guo, Y. Liang, P. Qin, Y. Yang, L. Luo, L. Leng, X. Gong and Z. Wu, *Small*, 2024, **20**, 2310396.
57 J. Min, J. H. Kim and J. Kang, *ACS Applied Nano Materials*, 2024, **7**, 26377–26396.
58 J. Yang, F. Bussolotti, H. Kawai and K. E. J. Goh, *physica status solidi (RRL)–Rapid Research Letters*, 2020, **14**, 2000248.
59 H. Cho, M. Sritharan, Y. Ju, P. Pujar, R. Dutta, W.-S. Jang, Y.-M. Kim, S. Hong, Y. Yoon and S. Kim, *Acs Nano*, 2023, **17**, 11279–11289.
60 Y. Liu, S. Guan, X. Du, Y. Chen, Y. Yang, K. Chen, Z. Zheng, X. Wang, X. Shen, C. Hu *et al.*, *Energy & Fuels*, 2023, **37**, 5370–5377.
61 T. Wahab, Y. Wang and A. Cammarata, *Physical Chemistry Chemical Physics*, 2023, **25**, 11169–11175.
62 A. Van De Walle, M. Asta and G. Ceder, *Calphad*, 2002, **26**, 539–553.
63 G. Kresse and J. Hafner, *Physical review B*, 1993, **47**, 558.
64 J. P. Perdew, K. Burke and M. Ernzerhof, *Physical review letters*, 1996, **77**, 3865.
65 J. Connolly and A. Williams, *Physical Review B*, 1983, **27**, 5169.
66 P. E. Blöchl, *Physical review B*, 1994, **50**, 17953.
67 S. Grimme, *Journal of computational chemistry*, 2006, **27**, 1787–1799.
68 J. Hafner, *Journal of computational chemistry*, 2008, **29**, 2044–2078.
69 D. J. Evans and B. Holian, *A)'=(A Is)*, 1985, **1**, 18.
70 A. Togo, L. Chaput, T. Tadano and I. Tanaka, *Journal of Physics: Condensed Matter*, 2023, **35**, 353001.
71 A. Togo, *Journal of the Physical Society of Japan*, 2023, **92**, 012001.
72 J. P. Perdew, K. Burke and M. Ernzerhof, *Phys. Rev. Lett.*, 1996, **77**, 3865–3868.
73 J. Heyd, G. E. Scuseria and M. Ernzerhof, *The Journal of chemical physics*, 2003, **118**, 8207–8215.
74 W. Setyawan and S. Curtarolo, *Comput. Mater. Sci.*, 2010, **49**, 299–312.
75 V. Zólyomi, N. Drummond and V. Fal'Ko, *Physical Review B*, 2014, **89**, 205416.
76 R. Pawar and A. A. Sangolkar, *Computational and Theoretical Chemistry*, 2021, **1205**, 113445.
77 T. Wahab, A. Cammarata and T. Polcar, *Phys. Chem. Chem. Phys.*, 2024, **26**, 29283–29297.
78 M. Fox, *Optical properties of solids*, Oxford university press, 2010, vol. 3.
79 J. S. Jang, H. G. Kim and J. S. Lee, *Catalysis today*, 2012, **185**, 270–277.
80 E. E. Benson, C. P. Kubiak, A. J. Sathrum and J. M. Smieja, *Chemical Society Reviews*, 2009, **38**, 89–99.
81 A. K. Singh, J. H. Montoya, J. M. Gregoire and K. A. Persson, *Nature communications*, 2019, **10**, 443.
82 K. Momma and F. Izumi, *Journal of Applied Crystallography*, 2011, **44**, 1272–1276.

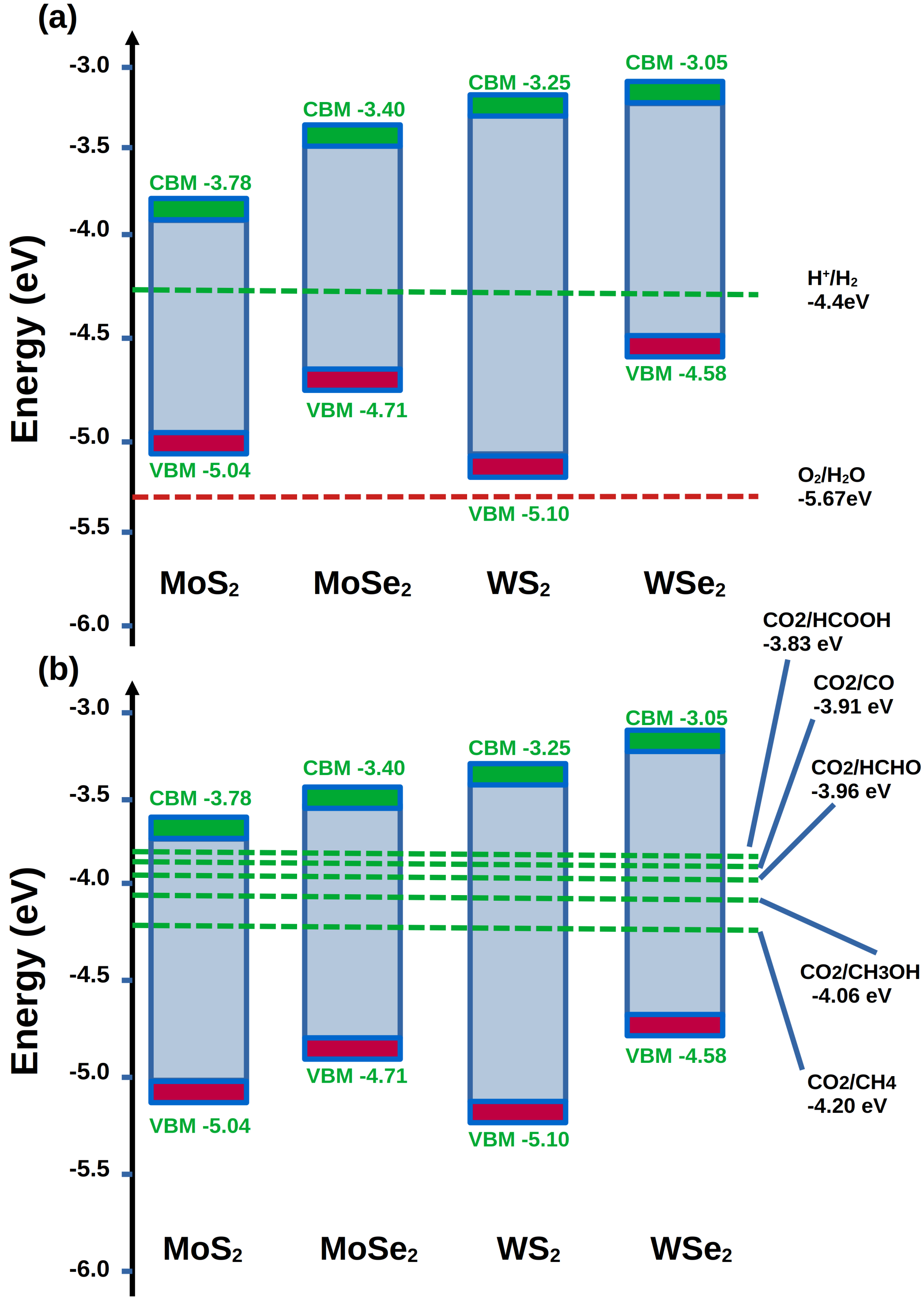

**Fig. 11** CBM and VBM of all $MX_2$ structure in comparison (a) water redox potential and (b) $CO_2$ reduction potential

# First principles study of chalcogen vacancy effect on the optoelectronic and photocatalytic properties of transition metal dichalcogenides monolayers — Electronic Supplementary Information

Tahir Wahab,*[a] Antonio Cammarata*[a] and Tomas Polcar [b]

## Contents

[a] Department of Control Engineering, Faculty of Electrical Engineering, Czech Technical University in Prague, Technicka 2, 16627 Prague 6, Czech Republic. Fax: +420 224 91 8646; Tel: +420 224 35 5713; E-mail: wahabtah@fel.cvut.cz, cammaant@fel.cvut.cz
[b] Engineering Materials & nCATS, FEE, University of Southampton, SO17 1BJ, Southampton, United Kingdom.

# 1 Optimised geometries

The optimized model geometries of the considered defective monolayers are provided in the following subsections.

## 1.1 $MoS_2$

```
MoS
1.00000000000000
12.6213897294682607   -0.0000259318422720    0.0000000000000000
3.1397230101016702   16.4168447338327681    0.0000000000000000
0.0000000000000000    0.0000000000000000   77.8784713744999948
Mo              S
24    45
Direct
0.6381915668048963    0.4472176258087801    0.1154978276738271
0.3879831622785905    0.4482202573929868    0.1151335312912963
0.7240408282357421    0.6102456057040295    0.1152536479419611
0.4707393065130828    0.6102917393707115    0.1151990787773246
0.8037002540097056    0.7759658923354937    0.1161001438843104
0.5553677600061502    0.7787748094299961    0.1157411889927530
0.8883780873257368    0.9430578417801769    0.1176912830783344
0.6357616284828275    0.9420596227607471    0.1164130401322844
0.1383441921989982    0.4470214968377440    0.1154267911009791
0.8878578330488175    0.4483483882739672    0.1153905199262023
0.2256026188924024    0.6096514353716914    0.1151714765099151
0.9698846614305978    0.6096260057748850    0.1152292805415441
0.3084224828078556    0.7759889572444961    0.1160717780369970
0.0560953289846799    0.7758489856821649    0.1169251570848756
0.3932394889697233    0.9420407947002846    0.1163852628027194
0.1401798210321549    0.9430188527964881    0.1176762527112316
0.4727304842042441    0.1093641986381338    0.1170339990663761
0.2229735002387054    0.1106167167436304    0.1176588618472190
0.5551834606481029    0.2790084372848006    0.1168773861609923
0.3055303206869491    0.2790234547431651    0.1168458713236001
0.9722337811758628    0.1109577050358843    0.1180224592319100
0.7217008686885324    0.1106170336970462    0.1177163518511617
0.0551811703852446    0.2791716466096952    0.1171483656812860
0.8052354794167405    0.2791665339015206    0.1171885263452892
0.5270846113123668    0.3875595719359404    0.0957546446964931
0.2793346346655497    0.3875758897187181    0.0957027720919306
0.6104282020238896    0.5584786738121331    0.0951826912509623
0.3614428284617079    0.5565661217875952    0.0949797814097430
0.5233170461410180    0.3969706758345462    0.1358260745089997
0.2780408941281317    0.3969350609392829    0.1357982721956320
0.6105974653768509    0.5563353823418938    0.1355339580660572
0.6919344766959473    0.7220449623917725    0.0955837941954466
0.4470997617589933    0.7221644673459756    0.0955545196160727
0.7774019477022155    0.8909487505898672    0.0967273987637031
0.5277432101163875    0.8893095475604299    0.0960221519565125
0.6938208371604905    0.7200808870371967    0.1358622156798116
0.4460809553153340    0.7200611305438778    0.1358431032092020
0.7707954854784714    0.8847083477265861    0.1368252058746254
```

```
0.0261747791022162    0.3872558931865285    0.0959269791444301
0.7797832211617468    0.3872313920982512    0.0959739280552287
0.1103148674294485    0.5585223029562529    0.0951296229120836
0.8603538470426085    0.5561997978725133    0.0951234962743317
0.0241984561632639    0.3973796642563908    0.1359945453605632
0.7772382521442851    0.3974746133887387    0.1360421586615634
0.1112788754071752    0.5561469051716247    0.1353859408897906
0.0287535831436759    0.8848738171986200    0.1376979940468379
0.1927216915402342    0.7227264404000494    0.0960312703238716
0.9457696563646113    0.7226740450113117    0.0960534991022182
0.2771445773457297    0.8909160076760371    0.0966991802161384
0.0275593267828217    0.8895119705767883    0.0974994738975313
0.2017762842582871    0.7157338802487682    0.1361348726313338
0.9404232224763349    0.7157413030422306    0.1361784916194940
0.2868162644485207    0.8847081951177276    0.1368057262771261
0.1949379021558241    0.2248766851621920    0.1373681825747615
0.3596560907348917    0.0547083452108542    0.0969907190331955
0.1104622988030401    0.0551541180918419    0.0978698979957750
0.4447835261564786    0.2214739104736222    0.0970120586606090
0.1939871743620730    0.2210172450621505    0.0972486406421703
0.3662404673415247    0.0503267679865160    0.1371367049433844
0.1127037859046699    0.0544338911590566    0.1379536795127324
0.4442422791978752    0.2229119064268757    0.1369800755511088
0.6925299171933968    0.2248715086381360    0.1374329644142069
0.8616652375624191    0.0551636634935937    0.0978986442791841
0.6130673849694405    0.0547437115508767    0.0970085648171348
0.9446546810999623    0.2207752377006795    0.0975340517214757
0.6955391676397497    0.2209487893709913    0.0972845929880397
0.8601119137974530    0.0544220021341886    0.1379920974108864
0.6083475706809844    0.0502462214735926    0.1371852050692307
0.9437546157855585    0.2251220173786499    0.1376130644640017
```

### 1.2 $MoSe_2$

```
MoSe
1.00000000000000
13.0828445326117890   -0.0000740622595253    0.0000000000000000
3.2742829995161151   17.0539676515872181    0.0000000000000000
0.0000000000000000    0.0000000000000000   80.0000000000000000
Mo               Se
24    45
Direct
0.6075266212671118    0.4719198226246434    0.2580376475542373
0.3681597944698114    0.4708915900203315    0.2582742875692700
0.6944799302636935    0.6372308009547576    0.2583472744765249
0.4492525769414048    0.6325303966350095    0.2581854974859700
0.7747904593862343    0.8048081294332687    0.2582139029001443
0.5350357196101433    0.8045365741637539    0.2580098413099396
0.8619766726587471    0.9683508159430971    0.2593115169028251
0.6141513157093680    0.9658193960952409    0.2581506844292047
0.1159761504330790    0.4684708947805480    0.2597427634638393
```

| | | |
|---|---|---|
| 0.8623777824689820 | 0.4697175908009962 | 0.2595845900698775 |
| 0.2001655233899039 | 0.6349717517008295 | 0.2594273871681328 |
| 0.9465074486484149 | 0.6362497619644905 | 0.2596747715593321 |
| 0.2836082650208853 | 0.8023022811797931 | 0.2596039507864538 |
| 0.0305414936518494 | 0.8028000743270778 | 0.2597999378947775 |
| 0.3665293246096286 | 0.9704829872052630 | 0.2584438516315959 |
| 0.1155345247158016 | 0.9696784614290133 | 0.2596800506641233 |
| 0.4418121662306360 | 0.1380027215012698 | 0.2580572263041538 |
| 0.2022349388094036 | 0.1382437822833254 | 0.2582534722990529 |
| 0.5273595615817047 | 0.3037885905198041 | 0.2584463134129353 |
| 0.2820749762990024 | 0.2990435371427070 | 0.2582290130271651 |
| 0.9474209749234582 | 0.1360334493084904 | 0.2597468938985720 |
| 0.6944868823476414 | 0.1354615135164810 | 0.2595518373208193 |
| 0.0326889156004502 | 0.3017040356107194 | 0.2593145539639832 |
| 0.7785707552621542 | 0.3029919032830103 | 0.2596100688394087 |
| 0.6428036491935276 | 0.3587824215468836 | 0.2378164841425567 |
| 0.3882702428920112 | 0.3598634611238569 | 0.2373989743235677 |
| 0.7254628770816158 | 0.5234860325206999 | 0.2377485575367301 |
| 0.4736714642495189 | 0.5254906734726433 | 0.2367130890979399 |
| 0.6328392528790325 | 0.3639683570889921 | 0.2796499460542770 |
| 0.3907894350226667 | 0.3585657495109630 | 0.2793727271297705 |
| 0.7145013752115212 | 0.5271174542912137 | 0.2796076478053480 |
| 0.8107575441348575 | 0.6922474125257725 | 0.2378571031639364 |
| 0.5554067019482252 | 0.6928210181821314 | 0.2372451418155975 |
| 0.8928346745529029 | 0.8573012345086720 | 0.2381760943169506 |
| 0.6421510509588997 | 0.8587856162235171 | 0.2366821346990183 |
| 0.7999909315472808 | 0.6963143058618687 | 0.2796877833753145 |
| 0.5577947389585628 | 0.6927718372268686 | 0.2792099909455938 |
| 0.8831716929417815 | 0.8574181552131470 | 0.2801304697507999 |
| 0.9747568060239414 | 0.5247253647793203 | 0.2807574523272754 |
| 0.1422177386085363 | 0.3589797333696717 | 0.2381879514424075 |
| 0.8910681122044668 | 0.3580875312568747 | 0.2387177555059847 |
| 0.2251089751825318 | 0.5241622098610330 | 0.2382477310226290 |
| 0.9758979311710083 | 0.5244973302199053 | 0.2388604898687474 |
| 0.1487665072942037 | 0.3523649275534229 | 0.2801790654817111 |
| 0.8922526312119017 | 0.3577039097018795 | 0.2804690665500426 |
| 0.2347027250686506 | 0.5237982246831718 | 0.2802801958387585 |
| 0.1443878885350037 | 0.8583717303960405 | 0.2808132492526280 |
| 0.3090881109139294 | 0.6922166098336251 | 0.2382065789849023 |
| 0.0587137238728173 | 0.6916880699376375 | 0.2388336911390885 |
| 0.3921723962068877 | 0.8569585746071382 | 0.2378175389077352 |
| 0.1426233128057993 | 0.8579379016081505 | 0.2388758042180014 |
| 0.3165342961048844 | 0.6863698300299923 | 0.2801687937120838 |
| 0.0601135686099496 | 0.6906912340527541 | 0.2806284834654088 |
| 0.4020813256404763 | 0.8601710101297882 | 0.2797324807677249 |
| 0.4773839061704696 | 0.0262929583436344 | 0.2373116625357697 |
| 0.2232817282265424 | 0.0257145358208358 | 0.2378995671903843 |
| 0.5583128979302165 | 0.1902671962219392 | 0.2378216070405317 |
| 0.3077094436915692 | 0.1922457501863613 | 0.2367060744385512 |
| 0.4758459303631853 | 0.0260750878117148 | 0.2792478813165732 |
| 0.2315396227061713 | 0.0299840065032077 | 0.2797808109827201 |

```
0.5474069377630898    0.1934186392361248    0.2797394035722225
0.8057847605188642    0.1915657067153465    0.2807376980411816
0.9756050778387947    0.0249990796891552    0.2387681009601618
0.7241868894921596    0.0254461940516709    0.2381118091920977
0.0580348036169570    0.1906504970986406    0.2381692013070889
0.8076920106278717    0.1911374004158201    0.2388351970933363
0.9740716628299416    0.0238720425061494    0.2805522909919991
0.7201654593063658    0.0195520682439520    0.2801018278354632
0.0671124035909484    0.1907540174139143    0.2801280029270347
```

### 1.3 $WS_2$

```
WS
1.00000000000000
22.9190387987332791    0.0001746768093215    0.0000000000000000
19.3939147253939872   12.2130421600155454    0.0000000000000000
0.0000000000000000    0.0000000000000000   80.0000000000000000
S                W
63    32
Direct
0.0492181476643486    0.5341621322388966    0.1331617870198417
0.4730928963818157    0.2308993396535041    0.1331059179357131
0.2308993047370883    0.4730928886625529    0.1331059686486036
0.2916237578828197    0.2916239804183076    0.0925602847211653
0.0419011300701079    0.5402854150919786    0.0939470514587715
0.4770276515844877    0.2308026387517758    0.0939100173671708
0.2308025408790316    0.4770277665673519    0.0939100708048253
0.6653521048503833    0.1667088724845747    0.1335543674664848
0.4161535171151112    0.4161535233274515    0.1336311255387171
0.8539009761654682    0.1039734714687173    0.1334582826932089
0.6040880554256438    0.3539859372256441    0.1335219097977258
0.6653425538138020    0.1670946316356008    0.0943486103129746
0.4163107085966875    0.4163109114132675    0.0944067327979460
0.8534823970641431    0.1041612830523746    0.0942918236590053
0.6036607950718919    0.3540762272083687    0.0943362891503482
0.0416608150632711    0.0416614362722312    0.1333918416976596
0.7917998081541479    0.2915875139408782    0.1334135608283055
0.2292083385322204    0.9796378928548656    0.1334275323101162
0.9796375229652355    0.2292086360334186    0.1334275517959843
0.0416346624938281    0.0416347849828281    0.0942297074457215
0.7916391753662739    0.2917243635878691    0.0942386479487579
0.2294807023754685    0.9795120106390496    0.0942509356655686
0.9795120585090314    0.2294807148254268    0.0942509680573640
0.1667085155302827    0.6653524481483533    0.1335544086332547
0.9162297297119192    0.9162303288468659    0.1334584980274021
0.3539858051784397    0.6040882226238578    0.1335219200619509
0.1039730036805400    0.8539014499120421    0.1334583007074431
0.1670946531674168    0.6653425134398714    0.0943486697537793
0.9162478781889773    0.9162476860755420    0.0943123788697501
0.3540761245890537    0.6036608698215062    0.0943363023475993
0.1041614143573116    0.8534822228698560    0.0942918582151160
```

```
0.5417374177572410    0.5417377913747690    0.1334192967316614
0.2915870530781620    0.7918002580059171    0.1334135842149553
0.7295510127184742    0.4791216364168773    0.1334674891968694
0.4791212802035100    0.7295514112367292    0.1334674923602035
0.5417199060115052    0.5417200409659808    0.0942496564419621
0.2917242063107968    0.7916392962432637    0.0942386801992743
0.7296609619975625    0.4794685225682691    0.0942900518651081
0.4794685058309538    0.7296609663844758    0.0942900461057123
0.9181766879098927    0.4162281710341607    0.1335879270359372
0.6668101472116391    0.6668105491554586    0.1336628838087892
0.1103518906302444    0.3522360175838030    0.1331277547410590
0.8554908976955437    0.6034370076945742    0.1337498245125641
0.9172461097645294    0.4172193308922130    0.0943692590003430
0.6671282436888745    0.6671281148228937    0.0944570028213561
0.1038732890563835    0.3558532817143827    0.0939269069303212
0.8542602487234007    0.6045095618086654    0.0945261745198906
0.4162278997550062    0.9181770015837747    0.1335879060846138
0.1676429838664575    0.1676430437549794    0.1333420947093015
0.6034365791294137    0.8554913779993755    0.1337497796002838
0.3522360507441716    0.1103518409215805    0.1331277038330443
0.4172194200771975    0.9172459396339432    0.0943692315929897
0.1670435361690757    0.1670435914944811    0.0942079979709968
0.6045097229718757    0.8542600655245451    0.0945261429684260
0.3558531238287158    0.1038733523514989    0.0939268513897335
0.7916825586480897    0.7916832899314479    0.1337129855264615
0.5341617015216715    0.0492186130123673    0.1331617299843246
0.9805626933434859    0.7266173515076404    0.1333831309015717
0.7266168287836953    0.9805632745158229    0.1333831023202080
0.7913819660786221    0.7913818247821404    0.0944981176274667
0.5402853747608986    0.0419011669041295    0.0939469825545464
0.9787108622995001    0.7289346384063030    0.0942444616697749
0.7289347503023814    0.9787107466908472    0.0942444140072698
0.1532196313667590    0.3907577194190847    0.1129619761558572
0.8971684038663921    0.6446959920538994    0.1140658075720276
0.3311417839984316    0.3311418064024079    0.1129300697463723
0.0849662458158725    0.5806827157190088    0.1135445021977462
0.5192899509910766    0.2710874526662100    0.1139508742708273
0.2710873953501118    0.5192899875689381    0.1139509113017438
0.7076520887285233    0.2082644321597195    0.1139373977904272
0.4580396204263421    0.4580396930028233    0.1139783124263333
0.8956713223905171    0.1457964554557117    0.1138205402384763
0.6457772346054804    0.3958066096507867    0.1138438933245749
0.0835165095622814    0.0835165545468608    0.1138094006447509
0.8336492263627101    0.3334185296835113    0.1138321414484074
0.0210012542132349    0.7698936358463980    0.1138957416449165
0.7698933901082875    0.0210014918128500    0.1138957194348436
0.2082643660346918    0.7076521716426096    0.1139374269627290
0.9580346052140749    0.9580348723169286    0.1138764188433136
0.3958063728444540    0.6457774558671475    0.1138439176150904
0.1457961594371140    0.8956716460159235    0.1138205806325951
0.5835715901033268    0.5835717984577021    0.1138431251409575
```

```
0.3334182842027653    0.8336494730234765    0.1138321592003350
0.7716347991135647    0.5208878375501882    0.1139866653127622
0.5208877175157823    0.7716349074691999    0.1139866597197917
0.9599129110501456    0.4580873524203163    0.1140034297930520
0.7086579016027081    0.7086579399607505    0.1141779251930502
0.2711786123446731    0.0214591796861075    0.1138756837730564
0.0214591203449390    0.2711786908838417    0.1138756911079865
0.4580872225331887    0.9599130014657942    0.1140033830880529
0.2093719355666808    0.2093720169287177    0.1135208508274161
0.6446957696052609    0.8971686810301598    0.1140657703946680
0.3907576665244589    0.1532196333712869    0.1129619155790337
0.8330455928779824    0.8330459254101998    0.1139736669698431
0.5806826472629597    0.0849663485193237    0.1135444346836498
```

### 1.4 $WSe_2$

```
1.00000000000000
13.1206362457420447    0.0014710624231543    0.0000000000000000
3.2782401790727471   17.0711920690310421    0.0000000000000000
0.0000000000000000    0.0000000000000000   80.0000000000000000
Se              W
45    24
Direct
0.5260111118542748    0.3887853233544605    0.1470875540479449
0.2804951171379100    0.3878441205831221    0.1462545809415732
0.6101512517864763    0.5567206401107949    0.1471589801992827
0.3626692203596217    0.5551121837183438    0.1443391255667206
0.5179395058296519    0.3950248275875556    0.1889767253075579
0.2787254361528331    0.3876676643816488    0.1879050445476648
0.5994849296606573    0.5570418489393516    0.1891509940966619
0.6930583839495732    0.7219842127656924    0.1471997775489625
0.4479903432935511    0.7207611565497947    0.1463972509057099
0.7768107479163042    0.8900942894898265    0.1472692225348904
0.5281334203522583    0.8882660663195833    0.1443784124392487
0.6857195576086205    0.7289281984422500    0.1890820409670624
0.4452509009830337    0.7220219182023304    0.1880651730238952
0.7677084191185958    0.8900260167462629    0.1890434417094702
0.8614254509219265    0.5558706153327162    0.1897095571270613
0.0258532523621050    0.3873829786339733    0.1468858991059975
0.7773176714986421    0.3887249927745221    0.1476873307068865
0.1117727824783408    0.5572533819471048    0.1467299477231727
0.8604099516511892    0.5559342730327419    0.1478678617798382
0.0369296775004939    0.3834562646809850    0.1886543085999941
0.7753458723306841    0.3899780773566009    0.1896549096589923
0.1230902935710870    0.5536062554984741    0.1885526267195292
0.0313130821796565    0.8889859460961934    0.1896324645980866
0.1939453430807492    0.7208410693576609    0.1467344450453786
0.9449274383472827    0.7224559812501504    0.1475701705971893
0.2786406538364097    0.8905225401329958    0.1466079584473551
0.0286541004244584    0.8894265294101649    0.1477841318065151
0.2043530888267712    0.7158195662994765    0.1885193190861408
```

0.9442212461212821 0.7229497724522950 0.1895161165880536
0.2902374478247463 0.8874078917025507 0.1886253915851938
0.3587381829400821 0.0543500918195786 0.1462693991536969
0.1125763220167996 0.0557002225282651 0.1470561139924965
0.4429653153396975 0.2238199441173839 0.1466185248835093
0.1942412539590694 0.2219151699354390 0.1443428207042262
0.3602430477830280 0.0544304374583200 0.1879056654886850
0.1174658263527650 0.0617496155312349 0.1889337334903210
0.4323627192400898 0.2206392362133604 0.1886333488865216
0.6907139348103076 0.2223347781251638 0.1897317166316708
0.8606006282352060 0.0555200101499569 0.1476438782428526
0.6128592150543057 0.0538118650335428 0.1468884232084021
0.9445088040406298 0.2234760181284364 0.1472632672200009
0.6932429180835079 0.2224404256238566 0.1478504433427303
0.8620960339838523 0.0567184662739627 0.1896390390750409
0.6037859911845281 0.0499006208721042 0.1887018185457981
0.9541489328617769 0.2232694500706827 0.1890425962231820
0.6355668834042337 0.4456490045644966 0.1682864523498774
0.3856743953733158 0.4522212696316164 0.1669260208522374
0.7193065386479203 0.6121908735012151 0.1687104113487428
0.4638742252393719 0.6075694783035994 0.1670241926048402
0.8034728337987016 0.7786771482498137 0.1682877658347822
0.5526456405555528 0.7857276794921259 0.1670336549299956
0.8884851224132722 0.9447188099612759 0.1687429749707608
0.6287475060483040 0.9398839329785259 0.1671561746771356
0.1413263486614121 0.4434411758565412 0.1672584881570094
0.8896796294290915 0.4436870589418013 0.1686673712050331
0.2346329746384886 0.6061710128405728 0.1668749954646570
0.9744689292926364 0.6110138750889926 0.1685603978994538
0.3085107853631789 0.7770608369749928 0.1672590124432105
0.0579665190865066 0.7769774365003292 0.1685282310459562
0.3990328182121907 0.9403965441822093 0.1667852330602325
0.1429443637621692 0.9452610889431388 0.1684290488907679
0.4697175177202651 0.1103557769616294 0.1672717258277776
0.2214310201419362 0.1189628159775616 0.1669072980547444
0.5511140984215626 0.2784313148304351 0.1684863728888808
0.2973956585909237 0.2738284819955734 0.1667554802618804
0.9742415925672658 0.1120200235986724 0.1682612753911408
0.7206664398347500 0.1102405455039421 0.1686645706365235
0.0676325233793015 0.2733618107133940 0.1671207271876926
0.8056688615728227 0.2778500003765962 0.1687865619154863